\documentclass[twocolumn,aip,apl,graphicx,unsortedaddress,10pt]{revtex4-1}
\usepackage{dcolumn}
\usepackage{amsmath,amssymb}
\usepackage{graphicx}

\newcommand{\fracd}[2]{\cfrac{\mathrm{d} #1}{\mathrm{d} #2}} 
\newcommand{\fracdd}[2]{\cfrac{\mathrm{d}^2 #1}{\mathrm{d} #2^2}}  

\begin{document}
\title{Enhancement of second harmonic generation in a doubly resonant metamaterial}
    \author{Tetsuo Kanazawa}
   \author{Yasuhiro Tamayama}
     \email{tama@giga.kuee.kyoto-u.ac.jp}
   	\author{Toshihiro Nakanishi}
   	\author{Masao Kitano}
   	  \affiliation{Department of Electronic Science and Engineering, Kyoto University, Kyoto 615-8510, Japan}
   	  \date{\today}
	  \begin{abstract}
	   We investigate second harmonic (SH) generation in a doubly
	   resonant metamaterial. 
	   We show that SH generation can be enhanced
	   when the resonant condition is satisfied for the SH frequency
	   as well as for the fundamental frequency. A unit cell of the
	   doubly resonant 
	   metamaterial consists of two coupled resonators, one of which
	   resonates at the 
	   fundamental frequency, whereas the other resonates around the
	   SH frequency. 
	   We observe that the SH generation in the doubly resonant
	   metamaterial is 4.6 times as 
	   large as that in a singly resonant metamaterial.
	  \end{abstract}
	  \maketitle
Metamaterials are arrays of artificial structures having dimensions that are
much smaller than the wavelength of electromagnetic waves. 
Metamaterials are used to realize media with unusual properties that 
may not be readily available in nature, e.g., negative refractive index 
media, transformation media, slow light media, and 
switchable media.\cite{zheludev}
The properties of
metamaterials are determined by the shape and material of the constituent,
and, therefore, devices with desired functions can be obtained by
designing the constituent. 

Pendry {\it et al.} showed that when a constituent of a metamaterial
is a resonant conductive element, electromagnetic energy is squeezed
into a small volume of the constituent at the resonant
frequency.\cite{pendry} They predicted that if a nonlinear material is placed
in the volume, the nonlinear behavior of the material is enhanced.
Thus far, this phenomenon has been applied to second harmonic (SH) and higher-order
harmonic generations,\cite{klein,klein07,shadrivov,kim,bias}
 nonlinear tunable metamaterials,\cite{shadrivov,wang,powell}
and bistable media.\cite{wang}

In this Letter, we introduce a doubly resonant metamaterial to generate
the SH waves more efficiently in the metamaterial than in the singly
resonant metamaterial used in the previous 
studies.\cite{klein,klein07,shadrivov,kim,bias} 
We demonstrate
that the SH generation can be enhanced if the resonant
condition is satisfied for the SH frequency as well as for 
the fundamental frequency. 
First, using an electric circuit model, we show that
SH generation can be further enhanced by introducing a resonator for 
the SH frequency in addition to 
a resonator for the fundamental frequency. 
Then, a unit structure of the metamaterial derived from
the circuit model is presented. Finally, experimental results for 
the SH generation using the microwave metamaterial are described.
Although the proposed method of enhancing the SH generation is
demonstrated in the microwave region, it can also be used in the
terahertz and optical regions. 

\begin{figure}[tb]
   \centering
   \includegraphics[scale=1]{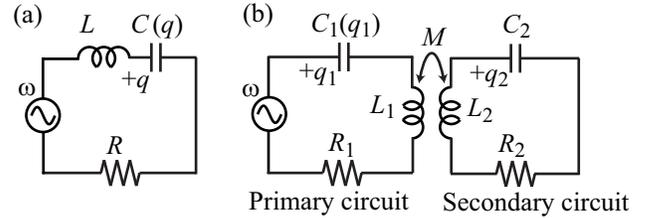}
   \caption{(a) Singly resonant circuit and (b) proposed doubly
   resonant circuit for enhancing the SH current. The circuit constants satisfy
   $1/\sqrt{L_1C_1}=\omega_1$ and $1/\sqrt{L_2C_2}=\omega_2\simeq 2\omega_1$.}
   \label{fig:figure1}
\end{figure}

We describe the concept of enhancing the SH generation using equivalent
 circuit models of resonant metamaterials. 
 A singly resonant metamaterial can be modeled as an 
 inductor-capacitor-resistor series resonant circuit, as 
 shown in Fig.~\ref{fig:figure1}(a). The external electromagnetic field
 corresponds to the voltage source $V\cos\omega t$.
 The voltage across the 
 capacitor reaches 
a maximum at the resonant angular frequency $\omega_0=1/\sqrt{LC}$. 
 If the capacitor exhibits nonlinearity, a large nonlinear effect 
 is expected at $\omega_0$.\cite{pendry}
 We assume that the voltage across the nonlinear capacitor
 is written as $v_{C}=q/C(q)=q/C+\alpha q^2$, where
 $q$ is the charge in the nonlinear capacitor, and $\alpha$ is the nonlinear
 coefficient. 
 Using the second-order perturbation method under a weak nonlinearity condition,\cite{poutrina} 
 the current amplitude oscillating at $2\omega$ is obtained as
 \begin{equation}
  \left|I(2\omega)\right|=\frac{\left|\alpha\right|V^2}{\omega^2\left|
   Z(2\omega)Z(\omega)^2\right|}\label{eq:1}, 
 \end{equation}
 where $Z(\omega)=R-\mathrm{i}\left[\omega L-1/\left(\omega C\right)\right]$ is the 
 impedance of the circuit. 
 When $\omega$ is tuned to $\omega_0$, 
 $\left|Z(\omega)\right|$
 takes a minimum value and $\left|I(2\omega)\right|$ 
 is maximized. 
 This implies that the strong SH signal is generated from the singly resonant
 metamaterial when $\omega=\omega_0$. Here, note that 
 $\left|I(2\omega)\right|$ is further enhanced by also reducing
 $\left|Z(2\omega)\right|$ in some way. That is, more efficient SH generation
 from the metamaterial can be achieved when the resonant conditions for
 both the fundamental and SH frequencies are satisfied. 

 We introduce a doubly resonant circuit shown in
 Fig.~\ref{fig:figure1}(b) for 
 the enhancement of the SH current. This circuit is composed of two
 resonant circuits: the
 primary circuit with resonant angular frequency
 $\omega_1=1/\sqrt{L_1C_1}$ and the secondary circuit with resonant
 angular frequency $\omega_2=1/\sqrt{L_2C_2}\simeq 2\omega_1$. These resonators
 are coupled via a mutual inductance $M$. 
  Kirchhoff's voltage law yields
  the coupled differential equations:
 \begin{align}
   L_1\fracdd{q_1}{t}+R_1\fracd{q_1}{t}+\cfrac{q_1}{C_1}+\alpha
   q_1^2-M\fracdd{q_2}{t}&=
   V\cos\omega t
   ,  \label{eq:2}\\
   L_2\fracdd{q_2}{t}+R_2\fracd{q_2}{t}+\cfrac{q_2}{C_2}-M\fracdd{q_1}{t}&=
   0.
  \label{eq:3}
 \end{align}
 Here, the voltage across the nonlinear capacitor
 is expanded up to the second
 order of the charge $q_1$.
 Applying the second-order perturbation method to Eqs.~\eqref{eq:2} and
 \eqref{eq:3}, we obtain 
 \begin{align}
  \left|I_1(2\omega)\right|&=
  \cfrac{\left|\alpha\right|
  V^2}{\omega^2\left|Z^\prime(2\omega)Z^\prime(\omega)^2\right|},  \label{eq:4}\\
  \left|I_2(2\omega)\right|&=\cfrac{\left|\alpha
  M\right|V^2}{\omega\left|
  Z_2(2\omega)Z^\prime(2\omega)Z^\prime(\omega)^2\right|},
  \label{eq:5}
 \end{align}
	where $Z_1(\omega)=R_1-\mathrm{i}\left[\omega L_1-1/\left(\omega C_1\right)\right]$ is the impedance of the primary circuit, 
	$Z_2(\omega)=R_2-\mathrm{i}\left[\omega L_2-1/\left(\omega C_2\right)\right]$ is
	the impedance of the secondary circuit, and 
	$Z^\prime(\omega)=Z_1(\omega)+\omega^2M^2/Z_2(\omega)$. 
	Equation~\eqref{eq:5} shows that, in order to maximize the SH current,
	$\omega_2$ must be shifted appropriately from $2\omega_1$, 
	considering $M$ and the finiteness of the quality factors. 
	Using a weak-coupling approximation, i.e., $\left|Z_1(\omega)Z_2(\omega)\right|\gg \omega^2M^2$ and $\left|Z_1(2\omega)Z_2(2\omega)\right|\gg 4\omega^2M^2$, 
	Eqs.~\eqref{eq:4} and \eqref{eq:5} can be simplified as follows:
	\begin{align}
	 \left|I_1(2\omega)\right|&\approx
	 \cfrac{\left|\alpha\right|V^2}{\omega^2\left|Z_1(2\omega)Z_1(\omega)^2\right|}
	 ,  \label{eq:6}\\
	 \left|I_2(2\omega)\right|&\approx\cfrac{\left|\alpha M\right|
	 V^2}{\omega{\left|Z_1(2\omega)Z_2(2\omega)Z_1(\omega)^2\right|}}.
	 \label{eq:7}
	\end{align}
	The SH current flowing in the primary circuit is
 	independent of $M$ and identical to 
	that in the case of the singly resonant circuit. 
	The SH current in the secondary circuit is proportional to $M$. 
	We define the current ratio as 
 	$\beta\equiv\left|I_2(2\omega)/I_1(2\omega)\right|\approx\left|\omega
 	M/Z_2(2\omega)\right|$. 
	When $\beta\gg 1$, the SH radiation power
	can be enhanced approximately by a factor of $\beta^2$
	compared with the case of the singly resonant metamaterial.

\begin{figure}[tb]
   \centering
   \includegraphics[scale=1]{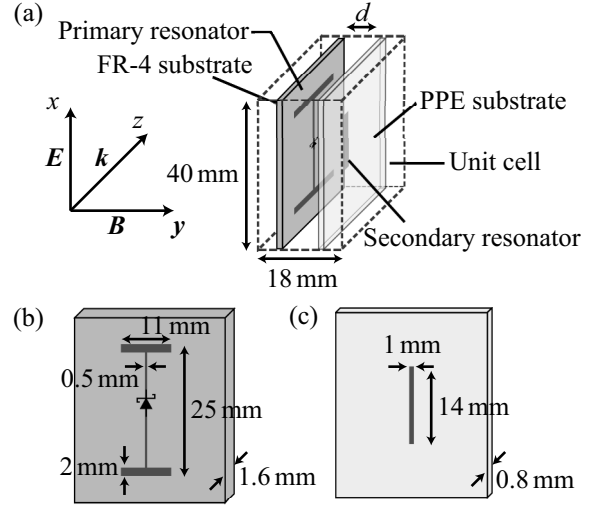}
   \caption{(a) Unit cell of doubly resonant metamaterial. The cells are
   aligned periodically in the  $x$ and $y$ directions.  Geometries of (b) the 
   primary resonator and (c) the secondary resonator. }
   \label{figure2}
\end{figure}

  Figure~\ref{figure2}(a) illustrates a unit cell of a
  doubly resonant metamaterial derived from the doubly resonant
  circuit. 
  The primary resonator shown in Fig.~\ref{figure2}(b) was
  fabricated on an FR-4 glass-epoxy printed circuit board of thickness
  $1.6\,\mathrm{mm}$ with a $35\,\mathrm{\mu m}$-thick copper layer.
  A Schottky diode (Rohm RB886G) was loaded at the center of the 
I-shaped structure and used as a nonlinear element as in the 
previous studies.\cite{shadrivov,wang,bias,powell}
  The periodic cascade of structures can be regarded as a series
  inductor-capacitor 
  resonant circuit.\cite{cle}
  The inductance is provided by a metal strip in the $x$ direction and
  the total capacitance is determined by the Schottky diode and gaps
  between the neighboring structures in the $x$ direction. 
  The primary resonator resonates at $3.4\,\mathrm{GHz}$.
  The secondary resonator shown in Fig.~\ref{figure2}(c) was
  made of a $35\,\mathrm{\mu m}$-thick copper film
  on a polyphenylene ether (PPE) substrate having a thickness of
  $0.8\,\mathrm{mm}$.
  We used the PPE substrate, which has a smaller dielectric loss than the FR-4 substrate, 
  to reduce the loss in the secondary resonator, i.e., to achieve the
  large enhancement factor of the SH generation. 
  The secondary resonator was designed to resonate at 
  $7.5\,\mathrm{GHz}$ considering $M$ and the quality factors. The two resonators were separated from each other by distance $d$. 
  Although the $x$-polarized incident electric field can directly 
  excite both the primary and secondary resonators, the current
  excited in the secondary resonator is negligibly small due to
  off-resonant excitation. 
  Therefore, the metamaterial can be modeled as the doubly 
  resonant circuit shown in Fig.~\ref{fig:figure1}(b). 

  For the experimental demonstration of the enhancement of the SH
  generation, we performed transmission
  measurements of the doubly resonant metamaterial. 
  The mono-layer doubly resonant metamaterial was placed in a parallel plate waveguide
  composed of two copper plates  parallel to the $yz$ plane with a separation of $40\,\mathrm{mm}$.
  The copper plates are equivalent to periodic boundaries 
  because the electromagnetic fields are uniform in the $x$ direction. 
  A pair of ultra-wide band dipole antennas\cite{diamond} was used as a
  transmitter and a receiver. The transmitting antenna was excited by a
  signal generator and 
  the receiving antenna was connected to a spectrum analyzer.

\begin{figure}[tb]
   \centering
   \includegraphics[scale=1]{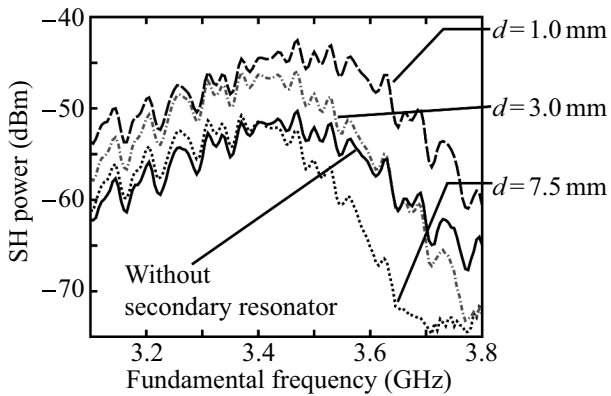}
   \caption{Measured SH power radiated from the doubly resonant
   metamaterial for three different $d$ and that from the
   metamaterial without the secondary resonator as a function of
   the fundamental frequency. 
   }
   \label{fig:figure3}
  \end{figure}

Figure~\ref{fig:figure3} shows the measured power of the SH wave
generated from the doubly resonant
metamaterial for three
different distances $d$. The SH power generated from the metamaterial without the
secondary resonator is also shown by the solid line. 
In the absence of the secondary resonator, the radiated SH power reaches a peak 
around the fundamental frequency of $3.4\,\mathrm{GHz}$,
which is the resonant frequency of the primary resonator. 
Focusing on the SH signal around the fundamental frequency
$3.4\,\mathrm{GHz}$, 
the observed SH power increases with decreasing $d$. 
The radiated SH power for $d=1.0\,\mathrm{mm}$ is 
4.6 times as large as that
in the case of the metamaterial without the secondary resonator.
The reason the doubly resonant metamaterial with large $d$ radiates less SH power
than the singly resonant metamaterial will be explained later. 

We compare the experimental results with the circuit analysis.
We assume that the SH generation from the secondary resonator is much
larger than that from the primary resonator. Based on Eq.~\eqref{eq:7}, the total SH radiation
power is written as
$
 P(2\omega)=aM^2\propto\left|I(2\omega)\right|^2, 
$
where $a$ is a proportional constant. 
 The mutual inductance $M$ between the primary and secondary
 resonators can be derived from Neumann's law\cite{stratton} and approximated as
 $M(d)\simeq
 \left(\mu_0/4\pi\right)\left(l_1+l_2\right)\log\left[\left(
 l_1+l_2\right)/d\,\right]$, where
  $\mu_0$
  is the permeability in a vacuum and $l_1\,(l_2)$ is
the effective length in the $x$ direction of the 
primary (secondary) resonator. We have assumed that the resonators are
infinitely thin conducting wires, 
that the current distribution is uniform along the $x$ direction, 
and $l_1, l_2\gg d$. We fitted the dependence of the measured SH power on $d$ to
$aM^2$ in the range of $d=2.5\,\mathrm{mm}$--$6.5\,\mathrm{mm}$ by varying
the parameters $l_1+l_2$ and $a$.

  \begin{figure}[tb]
   \centering
   \includegraphics[scale=1]{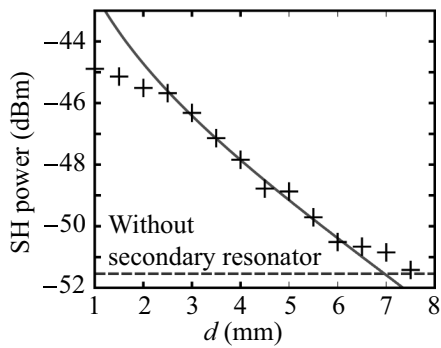}
   \caption{Measured (cross) and fitted (solid curve) SH powers versus
   $d$ at the 
   fundamental frequency $3.4\,\mathrm{GHz}$. The dashed line represents
   the SH power in the 
   absence of the secondary resonator. }
   \label{figure4}
  \end{figure}

Figure~\ref{figure4} shows the variation of the SH power for various $d$. 
The experimentally obtained SH power and the fitted curve agree well with
each other for $2.5\,\mathrm{mm}\le d\le 6.5\,\mathrm{mm}$.
Thus, we conclude that the SH generation is resonantly enhanced in 
 the secondary resonator coupled with the primary resonator via the
 mutual inductance. The reason the agreement of the measured values
 with the fitted curve is poor for $d<2.5\,\mathrm{mm}$ or 
 $d>6.5\,\mathrm{mm}$ is as follows. 
   When $d<2.5\,\mathrm{mm}$, $M$ becomes relatively large 
   and the weak-coupling approximation used in deriving
   Eq.~\eqref{eq:7}
   is not applicable. 
   In the case of $d>6.5\,\mathrm{mm}$, the SH wave radiated from the
   secondary   resonator is not large enough to neglect that from the primary
   resonator. 
   
For large $d$, the doubly resonant metamaterial radiates less SH power than
that from the singly resonant metamaterial at around the fundamental
frequency of $3.7\,\mathrm{GHz}$. In this case, the SH power
is radiated primarily from the primary resonator.
$\left|I_1(2\omega)\right|$ in Eq.~\eqref{eq:4} becomes smaller than
$\left|I(2\omega)\right|$
 in Eq.~\eqref{eq:1} because $\left|Z^\prime(2\omega)\right|>\left|Z(2\omega)\right|$. 
Thus, the total radiated SH power is smaller than that from
the singly resonant metamaterial. 

  We investigated the SH generation in the doubly resonant
  metamaterial. Using the electric circuit model, we showed that
  the efficiency of the SH generation can be enhanced when the resonant
  conditions for both the SH frequency and the fundamental
  frequency are satisfied. 
  We measured the power of the radiated SH wave
  from the microwave metamaterial and verified that the efficiency of the SH generation
  was further enhanced  by introducing the secondary resonator due to the
  resonance effect for the SH frequency.

The enhancement factor can be increased by reducing the losses in the
secondary resonator, increasing the coupling between the two resonators,
and finely tuning the resonant frequency of the secondary resonator; 
therefore, we expect larger SH radiation 
by optimizing the structure of the metamaterial.
The current method can be applied to other frequency-conversion
processes, such as higher-order harmonic generations and sum-frequency 
mixing. When using the proposed method in a multi-layered metamaterial,
it is necessary to design the metamaterial so that the phase-matching
condition is also satisfied.\cite{gorkunov}

The present research was supported in part by Grants-in-Aid for Scientific Research
Nos.\ 22109004 and 22560041, by the Global COE program
\textquotedblleft Photonics and Electronics Science and
Engineering\textquotedblright\ of Kyoto University, and by a research
granted from The Murata Science Foundation. One of the authors
(Y.T.) would like to acknowledge the support of a Research Fellowship of
the Japan Society for the Promotion of Science for Young Scientists.

\bibliographystyle{aipnum4-1}

\end{document}